\documentstyle[preprint,aps,epsf]{revtex}
\begin{document}
\draft

%%%%%%%%%%%%%%%%%%%%%%%%%%%%%%%%%
\title{Four Dimensional Quantum Topology Changes of Spacetimes}
%%%%%%%%%%%%%%%%%%%%%%%%%%%%%%%%
\author{Shuxue Ding${}^{\dagger}$, Yasushige Maeda${}^{\ddagger}$ and Masaru 
Siino\footnote{e-mail:msiino@th.phys.titech.ac.jp, 
JSPS fellow}\\
 ${}^{\dagger}$Department of Physics, Tokyo Institute of Technology\\ 
Meguroku, Tokyo 152, Japan\\
${}^{\ddagger}$Department of Mathematics, Faculty of Science and Technology,\\
Sophia University, 7-1 Kioi-cho, Chiyodaku, Tokyo 102, Japan\\
${}^{*}$Department of Physics, Kyoto University\\
Kitashirakawa, Kyoto 606-01, Japan} 
\author{Revised Version}
%%%%%%%%%%%%%%%%%%%%%%%%%%%%%

\maketitle
\begin{abstract}
We investigate topology changing processes in the WKB approximation of 
 four dimensional quantum cosmology with a negative cosmological
constant. As Riemannian
manifolds which describe quantum tunnelings of spacetime we consider 
constant negative curvature solutions of the Einstein equation i.e. hyperbolic
 geometries. Using four dimensional polytopes, we can explicitly construct
hyperbolic manifolds with topologically non-trivial boundaries
which describe topology changes. These instanton-like solutions are constructed out of 8-cell's,
16-cell's or 24-cell's and have several points at infinity called cusps.
The hyperbolic manifolds are non-compact because of the cusps but have
finite volumes. Then we evaluate topology change amplitudes in the
WKB approximation in terms of the volumes of these manifolds. We find
that the more complicated are the topology changes, the more likely are
suppressed.

\end{abstract}
%\end{titlepage}
%%%%%%%%%%%%%%%%%%%%%%%%%%%%%%%%%%%%%%%%%%%%%%%%%%%%%%%%%%%%%%%%
\section{Introduction}
In classical gravity the topology change of the universe can be considered only in the case that normally assumed principle like causality is violated\cite{GST}. In quantum gravity, on the other hand, topology 
changing processes are rather ubiquitous. For example, 
topology changes can happen in the birth of the universe, in the evaporation
of a black hole and so on. Moreover, the topology changes, if 
possible by any mechanism either in classical or quantum gravity, may induce important physical effects. 

Recently there have been new progresses in the investigation of topology changes.
In (2+1)-dimensional simplified models, the topology changes have been
demonstrated to happen indeed by some explicit examples. In
three dimensional spacetimes with a negative cosmological constant, two
kinds of topology changes have been investigated. The first one was
associated with the existence of a compactified three dimensional
black hole solution (or a higher genus universe with a negative
cosmological constant). One of present authors (M. S.) showed that its
analytical continuation around the coordinate singularity of the
spacetime may provide a process of topology change\cite{MS}. The
second one was in the context of quantum cosmology. Fujiwara, Higuchi,
Hosoya, Mishima and one of the present authors (M.~S.)\cite{FHHMS}
constructed topology changing solutions by quantum
tunneling. To discuss more physical topology changes these works should
be generalized to four dimensional spacetimes. The former will be
generalized to the (3+1)-dimensional compact hyperbolic
cosmology\cite{IKS}. The purpose of the present article is the
generalization of the latter. That is to say, we investigate quantum topology changes
 in a cosmological model through tunneling
processes in four dimensions.   

According to Gibbons and Hartle\cite{GH}, a quantum 
tunneling spacetime is semi-classically approximated by
a Riemannian manifold with totally geodesic boundaries. In 
Ref.\cite{FHHMS}, the authors found some constantly curved
Riemannian manifolds with topologically non-trivial, totally geodesic
boundaries. If such a manifold has two connected pieces of boundary components with
different topologies, it describes a process of topology change
between the two Lorentzian spacetimes connected on the boundaries. Such
3-manifolds were constructed out of regular truncated polyhedra embedded
in a hyperbolic 3-space. In the present article, we discuss the topology change
 of the universe along the same line 
but in four dimensions. The four dimensional analogue of polyhedra is called polytope and we shall construct four dimensional Riemannian
manifolds by four dimensional regular truncated polytopes embedded in
a hyperbolic 4-space\cite{TH}. The resultant manifolds describe
topology change of a vacuum universe with a negative cosmological 
constant. 

In the next section I$\!$I we briefly review quantum tunnelings of
spacetimes in general, and give a mathematical preliminary for hyperbolic
space and quotient manifolds. The section I$\!$I$\!$I gives topology 
changing solutions in four dimensional spacetimes of a constant negative curvature. We
investigate their amplitudes and discuss the strong rigidity of the tunneling manifolds in the section I$\!$V. The
final section V is devoted to summary and discussions. 
%%%%%%%%%%%%%%%%%%%%%%%%%%%%%%%%%%%%%%%%%%%%%%%%%%%
\section{Quantum Tunneling of Spacetimes and Hyperbolic Manifolds}
\subsection{Quantum Tunneling of Spacetimes --- formalism}
In the context of quantum cosmology, a quantum tunneling should be
described by Riemannian path integral formalism proposed by Hartle
and Hawking \cite{HH}. We would like to appeal to the WKB approximation to compute tunneling amplitudes, since exact computations are almost hopeless in four dimensions. In this case, a quantum
tunneling means a transition (classically forbidden)
from a spatial hypersurface ${\Sigma}_i$ to another 
spatial hypersurface ${\Sigma}_f$. By topology
change of spacetime it is meant that ${\Sigma}_f$ is topologically
different from ${\Sigma}_i$. These hypersurfaces may consist of
some disconnected components. 

\begin{figure}
	\centerline{ \epsfbox{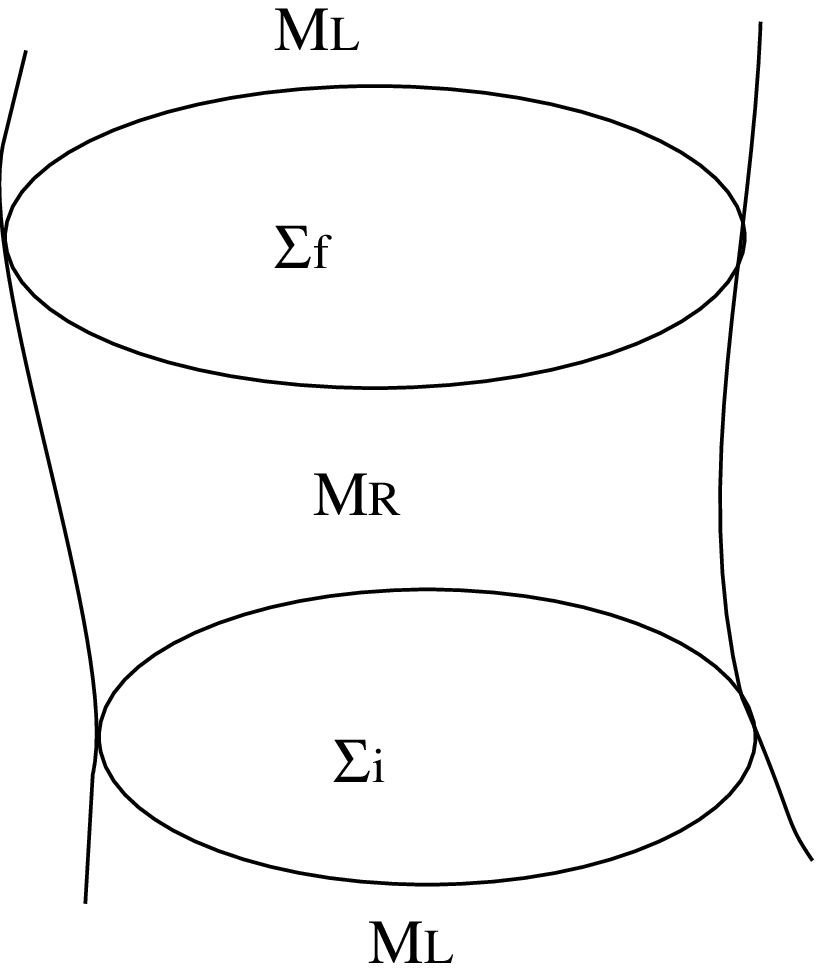}}
        \caption{A manifold with a Euclidean signature $M_R$
         interpolates the two manifolds with a Lorentz signature
         $M_L$ through $\Sigma_{i,f}$}
        \protect\label{fig:tun}
\end{figure}

Gibbons and Hartle \cite{GH} showed that, in the WKB approximation,
the tunneling process is described by a Riemannian manifold which has
the boundary components $\Sigma_i$ and $\Sigma_f$. In the ADM-formalism, a
spatial hypersurface $\Sigma$ is characterized by a spatial metric $h_{ij}$ 
and an extrinsic curvature $K_{ij}$ on it. In a semi-classical
picture, an ordinary spacetime manifold $M_L$ with a Lorentzian
signature (Lorentzian manifold) and a quantum tunneling manifold $M_R$ with a Euclidean
signature (Riemannian manifold) are connected on the hypersurface 
$\Sigma$ (see Fig.\ref{fig:tun}).
The spatial metric $h_{ij}$ can be uniquely defined in the viewpoints 
of both of regions, because it is independent of the 
time coordinates. However, the hypersurfaces connecting $M_{L}$ and $M_{R}$ cannot be arbitrarily chosen. Now we work out 
the condition the connecting hypersurfaces should satisfy. By a lapse function $N$ and a shift
vector $N^i$, the extrinsic curvature of $\Sigma$ is defined as
\begin{equation}
        K_{ij} = -{1 \over 2N}\left({\partial h_{ij}\over \partial 
        \tau}-D_{(i}N_{j)}\right),
%        \label{}
\end{equation}
in the Riemannian manifold, where $\tau$ is the time coordinate 
in this region, and it is defined as
\begin{equation}
        K_{ij} = -{1 \over 2N}\left({\partial h_{ij}\over \partial
        t}-D_{(i}N_{j)}\right),
%        \label{}
\end{equation}
in the Lorentzian manifold, where $t$ is the time coordinate in
this region. In these definitions, $D_i$ is the covariant derivative 
with respect to $h_{ij}$. Since the time $\tau$ in $M_R$ is analytically 
continued to the time  $t$ in $M_L$ as $\tau=it$ at $\Sigma$,
 the analytical continuation of geometrical variables, $h_{ij}$ and 
$K_{ij}$, requires vanishing $K_{ij}$ and $N_i$ at $\Sigma$.
Hereafter the boundary hypersurfaces with vanishing 
extrinsic curvature will be called totally geodesic boundaries. Then
on the connecting hypersurface, it must be totally geodesic.

For the sake of cosmological interest and simplicity, in the present
article, we consider a vacuum spacetime with a cosmological constant. If we further assume vanishing Weyl curvature, the spacetime has a geometry with constant curvature. 
Thus the Riemannian tunneling manifold $M_R$ becomes locally 
isometric to one of the following cases, $S^4$ (4-sphere), $R^4$ (4-plane) or 
$H^4$ (4-hyperboloid). In Ref.\cite{GH}, however, it was proved that 
if a 4-manifold has two pieces of disconnected boundaries $\Sigma_i$
and $\Sigma_f$, the spacetime should violate the energy condition at
some points. The energy condition states  
\begin{equation}
        R_{\mu \nu}V^{\mu}V^{\nu}>0
        \label{enc}
\end{equation}
for all vector $V^{\mu}$. Therefore we can exclude $S^4$ from our
considerations of topology changing manifolds because the
curvature of it is positive definite.  

From the Gauss-Codazzi equation, the vanishing extrinsic curvature 
makes $\Sigma_i$ and $\Sigma_f$ also have constant curvatures 
(locally isometric to $S^3, R^3$ or $H^3$) if the 4-manifold has a
constant curvature. To consider topology changes we need 
variety of the topologies. The Riemannian manifold
locally isometric to $R^4$ does not satisfy this requirement because the
topology of $\Sigma$ is too restricted (see \cite{KHT}).
 On the other hand, since the variety of hyperbolic 3-manifolds
(Riemannian manifolds locally isometric to  
$H^3$) is very rich, we will consider the Riemannian manifolds which
is locally isometric to $H^4$. A vacuum spacetime with a negative 
cosmological constant can just serve our purpose. Then the main question 
we want to answer in the present article is posed as 
\begin{quote}
Can we construct a hyperbolic 4-manifold with totally geodesic 
boundaries $\Sigma_i$ and $\Sigma_f$ which have different topologies?
\end{quote}

Mathematically, hyperbolic 4-manifolds are
quotient manifolds of a 4-hyperboloid $H^4$ by discrete subgroups
of its isometry group $SO(4,1)$. The fundamental region of this
quotient 4-manifold is a 4-polytope (four dimensional objects bounded
 by a collection of
polyhedra) embedded into $H^4$. Intuitively, a quotient manifold
means taking some copies of the fundamental regions with their
faces identified pairwisely. If we need a 4-manifold with
boundaries, some of the 3-faces of the fundamental regions should
remain  unidentified which form the 3-boundaries of the
4-manifold. Previously, some (2+1)-dimensional analogues of this have
been discussed\cite{FHHMS}. Although it is more complicated in 
four dimensions, by similar procedures, we can determine the fundamental
region and then the identifications of its 3-faces in hyperbolic
geometry \cite{TH}.  
%%%%%%%%%%%%%%%%%%%%%%%%%%%%%%%%%%%%%%%%%%%%%%%%%%%%%%%%%%%

\subsection{Hyperbolic Geometry and Klein Model}
To give a hyperbolic structure of four dimensional polytopes we embed
them into $H^4$. In our constructions, we use
$n$-dimensional Klein model (projective model)\cite{TH}
as the model of hyperbolic geometry.
The $n$-Klein model is a model on an open $n$-disk
\begin{equation}
        D^n=\{x^i\in R^n\vert x^i x_i<1\},
        \label{disk}
\end{equation}
in which a metric is
\begin{equation}
        ds^2 = {1 \over 1-r^2}\left({dr^2 \over 1-r^2} +r^2 
        d\Omega_{n-1}^2\right).
        \label{metric}
\end{equation}
As $r$ goes to $1$, one approaches a sphere at infinity 
$\partial D^n$. This metric gives a constant sectional curvature 
$-1$ and has a hyperbolic structure. Then this Klein-model is 
isometric to the spatial hypersurface of the well-known 
$n$-dimensional open-universe ($k=-1$).

\begin{figure}[t]
	\centerline{ \epsfbox{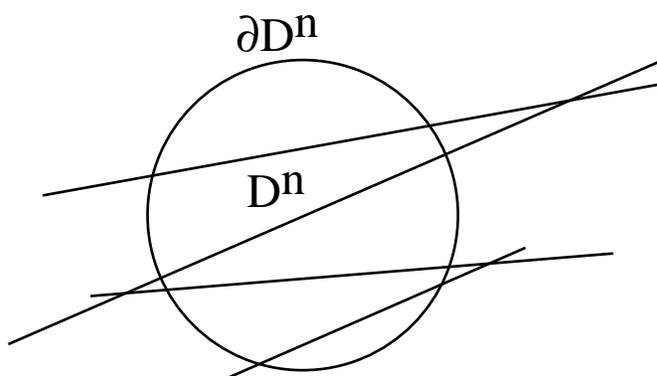}}
        \caption{A two dimensional example. There are some lines
         not intersecting each other inside the Klein model $D^n$. 
         They are parallel lines and intersect outside the sphere 
         at infinity $\partial D^n$. }
        \protect\label{fig:klm}
\end{figure}

\begin{figure}[t] 
	\centerline{ \epsfbox{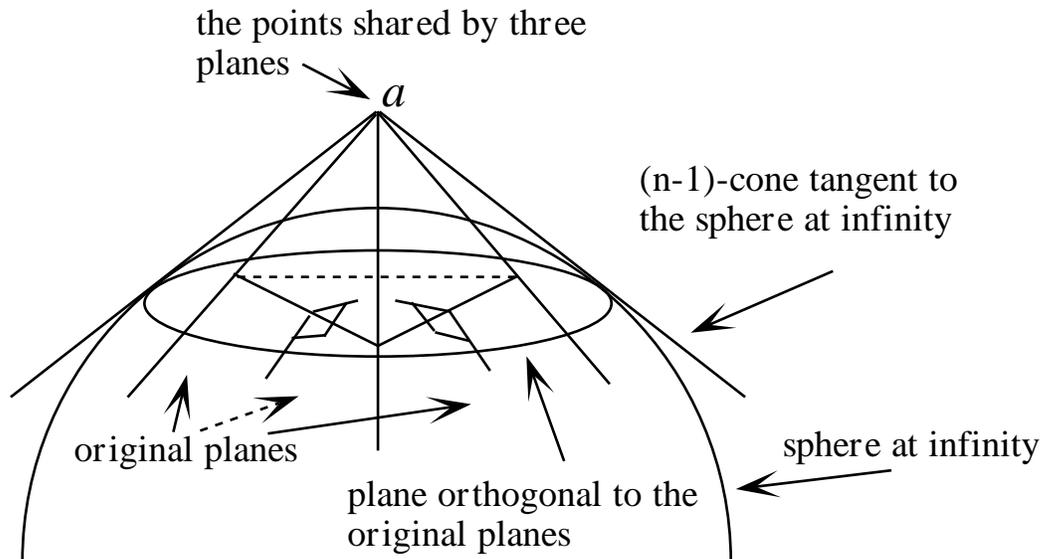}}
        \caption{Three planes share the point `$a$' outside the sphere. The cone
        with the vertex  `$a$', is tangent to the sphere at infinity. 
        Then the plane through the tangent points is perpendicular 
        to the three planes. }
        \protect\label{fig:poi}
\end{figure}

Here we briefly review the important properties of this model. First,
it is easy to find that all totally geodesic (extrinsic curvature
vanishing) $m(<n)$-hypersurfaces are $m$-planes in this model. Then we can 
construct totally geodesic boundaries by connecting such $m$-planes.
For example, if $m=n-1$, the $m$-planes bound a polytope. 
We know that each $m$-plane can be identified with another 
by the isometry $SO(n,1)$ of $H^n$. By these identification we shall
construct quotient manifolds. A more important property 
arises when we consider ideal points outside the sphere at infinity 
$\partial D^n$.  As depicted in Fig.\ref{fig:klm}, most of `parallel' 
$(n$$-$$1)$-planes, which do not intersect each other inside the Klein
model $D^n$ but intersect outside the sphere at infinity $\partial D^n$. 
For our purpose we consider situations in which some
$(n$$-$$1)$-planes share only one point `$a$' outside the sphere at 
infinity. The
$(n$$-$$1)$-planes form a pyramid with the vertex `$a$'.
There ought to exist a special cone which is tangent to the
sphere at infinity. There exists an $(n$$-$$1)$-plane which intersects the cone at the tangent points (exemplified in Fig.\ref{fig:poi}). The virtue 
of the Klein model is that this $(n$$-$$1)$-plane is 
orthogonal to all of the planes forming the pyramid.

\begin{figure}[t] 
	\centerline{ \epsfbox{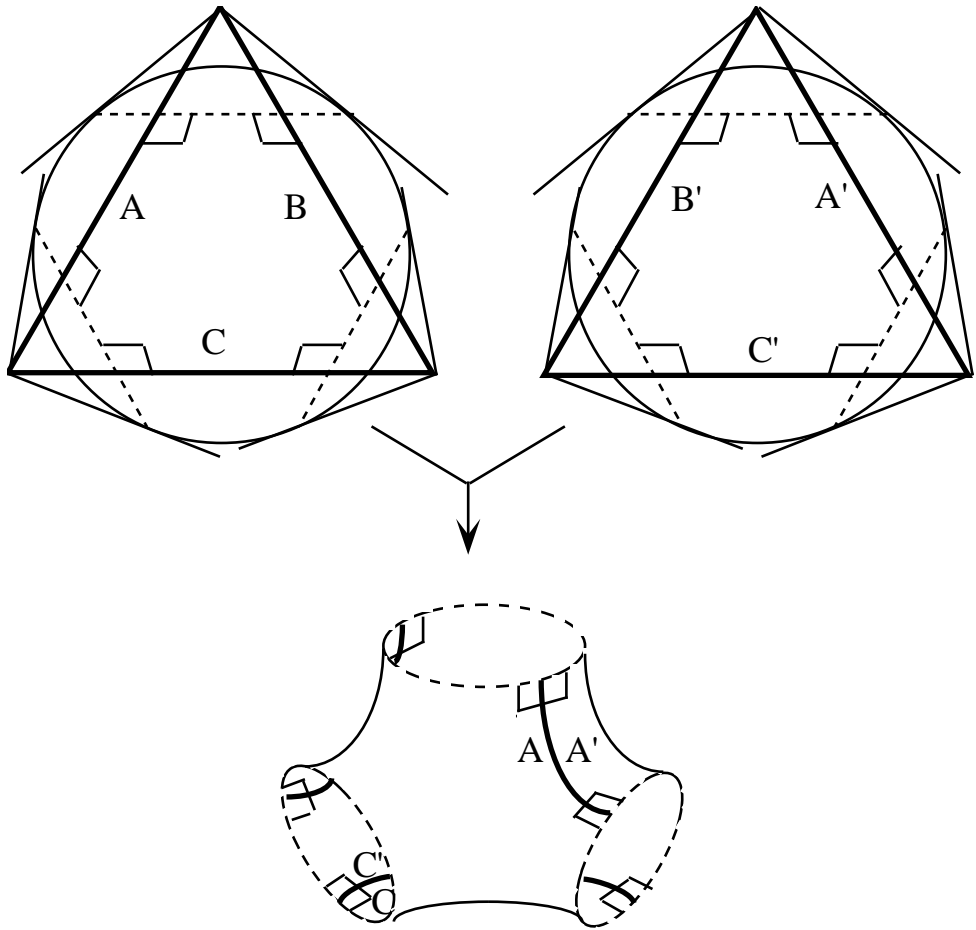}}
        \caption{Two hexagons made by regular truncations of triangles 
         are glued. The resultant space is one of the simplest topology 
         changing manifolds.}
        \protect\label{fig:ex2}
\end{figure}

These facts facilitate our procedure of construction of tunneling manifolds. As an example, we
show a simple case of a 2-Klein model of $H^2$. In
Fig.\ref{fig:ex2}, two regular triangles are drawn in the 2-Klein
model so that each vertex protrudes from $D^2$.
As mentioned above, we can draw dotted lines which are orthogonal to
the edges of the triangle and truncate off the vertices along these
dotted lines. We call such a truncation a regular truncation.
Gluing the triangles so that the labeled edges match each other, we get
a hyperbolic manifold with three $S^1$ boundaries. Since the lines
composing the boundaries are geodesic and orthogonal to the edges 
of the regular triangle, the boundaries become smooth totally geodesic
$S^1$\cite{DB}.

Finally we should remark upon the relation between the size of 
an object bounded by planes and the angles between these 
planes. In the hyperbolic geometry, if we enlarge the size of the object, 
the angles decrease. When the size approaches zero, the angles 
become the same values in Euclidean geometry. An angle on the
sphere at infinity $\partial D^n$ vanishes. We
have no well-defined angle outside the sphere at infinity $\partial  
D^n$.

%%%%%%%%%%%%%%%%%%%%%%%%%%%%%%%%%%%%%%%%%%%%%%%%%%%%%%%%%%
\subsection{Four Dimensional Polytopes}
First we prepare regular truncated 4-polytopes in the 4-Klein model of 
hyperbolic 4-space $H^4$. Regular 4-polytopes are 5-cell, 8-cell, 16-cell, 
24-cell, 120-cell and 600-cell\cite{COX}. For example, ``5-cell'' means
 that there are five congruent
polyhedra which bound the 4-polytope. We shall consider large polytopes in the
4-Klein model in the following sense:
\begin{description}
        \item[1)] All the vertices are outside of the sphere at 
        infinity.
        
        \item[2)]  Each edge of polytopes has intersections with 
        the sphere at infinity.
\end{description}
The above conditions guarantee that a single ideal vertex shared by 3-planes
which are cells (polyhedra) bounding the polytope can be regularly truncated off. 
As a generalization of the discussion in the last subsection, to a vertex,
there is a unique 3-plane which is perpendicular to the polyhedra 
bounding the polytope. Also as mentioned in the previous subsection, the 
dicellular angle (the angle between two adjacent polyhedra in a 
4-dimensional space) decreases as the size of the polytope increases 
in the hyperbolic geometry. To produce a regular and smooth structure 
after gluing of polytopes, we choose the size of the polytope to make
dicellular angles become $2\pi/n$  
($n$ is an integer). Then the size of the polytopes are restricted 
further to some discrete values. From a geometrical calculation we can
find the allowed dicellular angles of the polytopes. The allowed
polytopes are shown in the table 1.

The first column gives the names of polytopes and the second column 
the polyhedra which bound the polytope. Allowed dicellular 
angles are shown in the third column. The fourth column shows polyhedra 
produced by regular truncations. The solid 
angles of these polyhedra at their vertices are in the fifth column. Here it should be noticed 
that the edges of the polytope are tangent to the sphere at infinity 
except for the cases of a 24-cell with a dicellular angle 
$2\pi/5$ and a 600-cell with a dicellular angle $2\pi/3$. 
Therefore, in most situations, vertices made by the 
truncation are on the sphere at infinity (This aspect is reflected in the 
fifth column since the vertices at infinity have vanishing solid angles).
In these cases, the truncated polytopes are of course non-compact
(exemplified for the case of 8-cell in the   
next section). Calculating the volume, however, in the next 
section, we find that their volumes are finite. In the present article, we 
only consider these non-compact cases. By allowing the points at 
infinity, the construction of the tunneling manifold becomes much easier. Because a constructed
object is required to be a manifold, we should 
consider completeness of the construction which gives a restriction
at every vertex generally. However, in the special cases with
vertices on the sphere at infinity, these restrictions do not exist.

\begin{figure}[t]
	\centerline{ \epsfbox{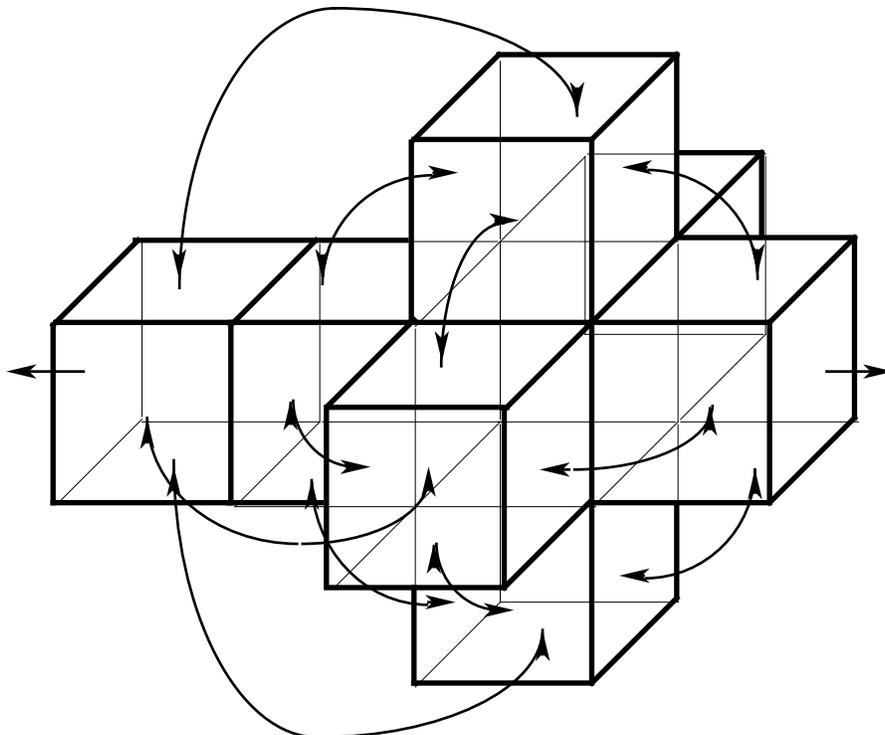}}
        \caption{The development of an 8-cell. Gluing the faces 
         of the hexahedra along the arrows in four 
         dimensions, we get the 8-cell of a 4-dimensional polytope.}
        \protect\label{fig:unf}
\end{figure}

In the next section we shall demonstrate some examples of
constructions of Riemannian manifolds which describe topology
changing processes. One of such Riemannian manifolds is constructed from
twelve 8-cell's which are 4-polytopes bounded by eight
congruent hexahedra. The development of such an 8-cell on 3-space is
shown in Fig.\ref{fig:unf}. Gluing faces in four dimensions according
to the arrows in Fig.\ref{fig:unf}, we get a 4-dimensional polytope
bounded by these eight hexahedra, which has sixteen vertices.
%%%%%%%%%%%%%%%%%%%%%%%%%%%%%%%%%%%%%%%%%%%%%%%%%%%%%%%%%%%%%%%%%%%%%%%%%%%%%%
\section{Riemannian Manifolds with Totally Geodesic Boundaries}
\subsection{Construction from 8-cell}
\begin{figure}[t]
	\centerline{ \epsfbox{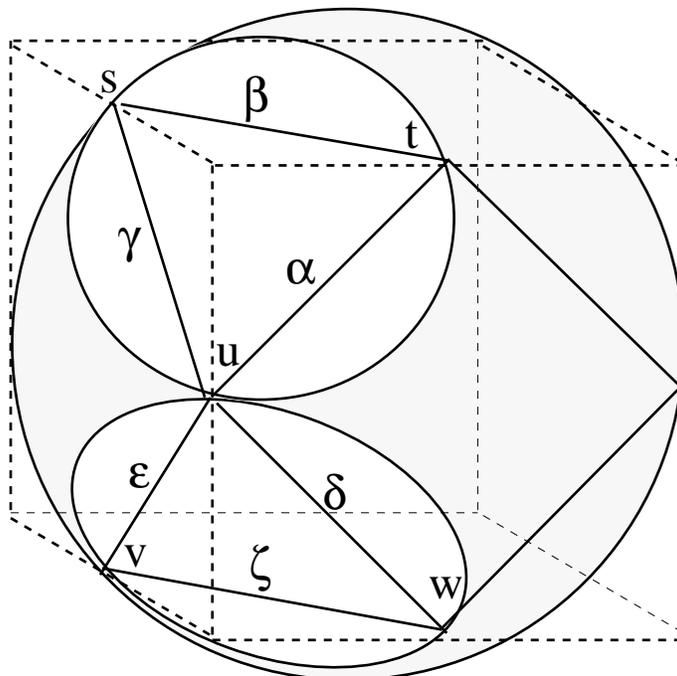}}
        \caption{The shaded sphere is a sphere at infinity in the 
        3-Klein model. Each edge of the hexahedra is tangent 
        to the sphere at $s,t,u...$. The sphere is cut by 
        planes through $s,t,u...$. Along these planes we 
        truncate the vertices of the hexahedron.}
        \protect\label{fig:cus2}
\end{figure}

We can adjust the size of the embedded 8-cell so that all the dicellular 
angles are $\pi/3$. Then we can show that all vertices are located outside 
the sphere at infinity $\partial D^4$. This 8-cell satisfies the two
conditions given in the previous section. The edges of the 8-cell are
coincidentally tangent to the sphere at infinity. Each hexahedron of
the 8-cell is embedded into an induced 3-Klein model (sub-model of the
4-Klein model) as shown in Fig.\ref{fig:cus2}. In this three
dimensional figure, all vertices are also outside the sphere at
infinity $\partial D^3$ and all edges are tangent to the sphere.

To get smooth totally geodesic boundary hypersurfaces, we truncate every
vertex of the 8-cell in an analogous way as we did in the 2-dimensional example in I$\!$IB.
Let us pay attention to the four hexahedra having a vertex in common
in Fig.\ref{fig:unf}. The property of the Klein model guarantees the
existence of a unique 3-hyperplane which is perpendicular to all of
the four hexahedra as mentioned in the previous section. In this way we
cut out the regions near the sixteen vertices of the 8-cell by these perpendicular
3-hyperplanes to get a regular 
truncated 8-cell embedded completely in the 4-Klein model. The 
truncation of 8-cell induces truncation on every hexahedron bounding
the 8-cell. The resultant hexahedron is shown in Fig.\ref{fig:cus2}. 
On each hexahedron the truncation of the vertex of the 8-cell makes 
a triangle with its vertices on the sphere at infinity $\partial
D^3$. It is noticed that the triangles share 
vertices with adjacent triangles (for example, the vertex (u) is shared by the two adjacent triangles $\triangle stu$ and $\triangle uvw$ in
Fig.\ref{fig:cus2}). In this case, any edge of the original hexahedron can be completely
truncated off by the two truncations of the two adjacent vertices connected by
 the edge (see Fig\ref{fig:cus2}).
 
\begin{figure}[t]
	\centerline{ \epsfbox{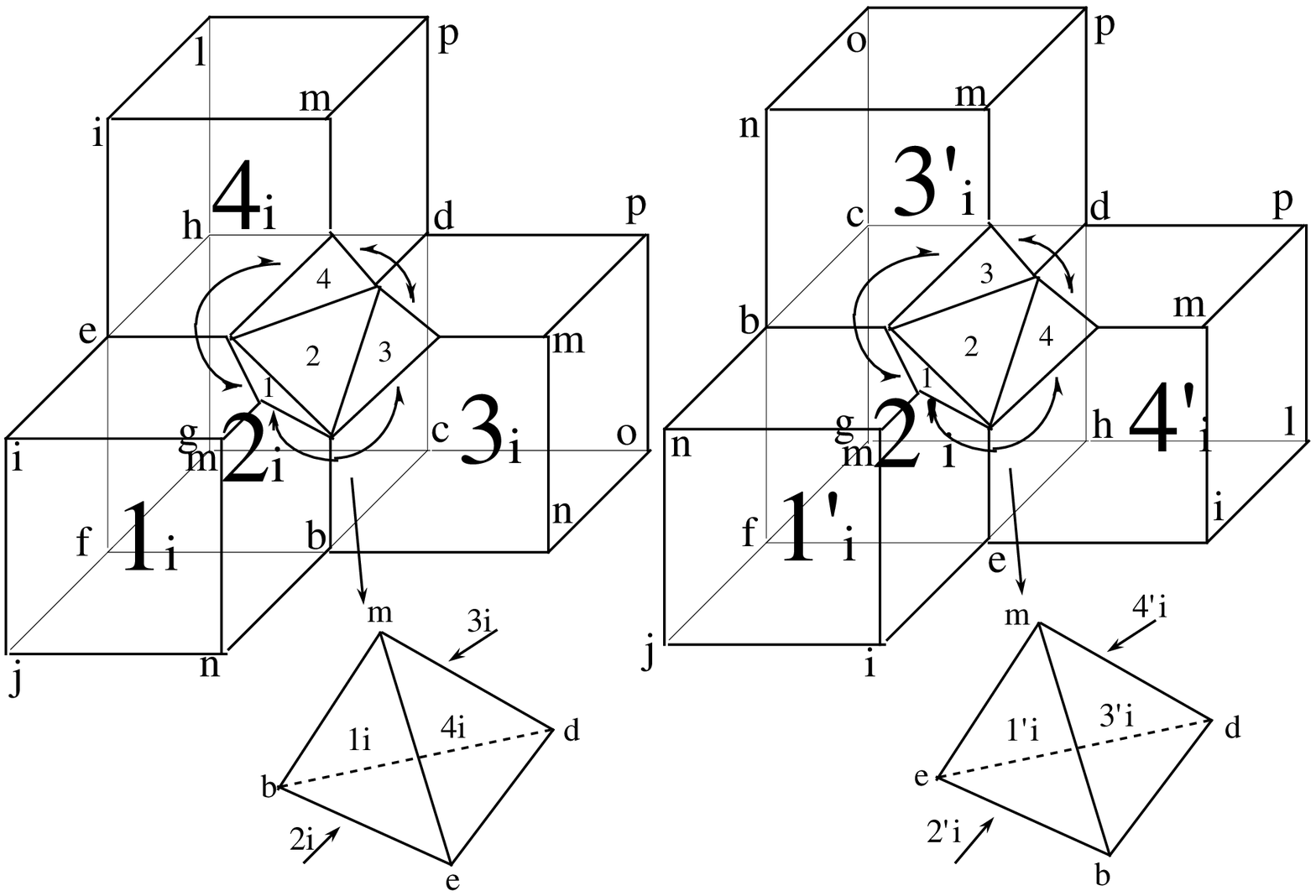}}
        \caption{A part of the 8-cell in Fig.5. Four hexahedra meet 
	at a vertex. When we truncate the
vertex, a tetrahedron appears. We label the faces and the
vertices of the tetrahedra at the vertices $(a)$ by the index number
of the cell which the face belongs to and the vertices of the 8-cell's
on the opposite side of $(a)$, respectively.}
\protect\label{fig:trt}
\end{figure}

  Because four hexahedra share one
vertex in an 8-cell 
(see Fig.\ref{fig:trt}), each 3-boundary of the 8-cell made by the
truncation of a vertex is bounded by four triangles. From Fig.\ref{fig:trt} we see
that the 3-boundary is a tetrahedron whose vertices are on the sphere
at infinity (To the cases of other 4-polytopes, see the fourth column of 
the table 1). Since each tetrahedron is orthogonal to the hexahedra in
the hyperbolic 4-space, the dihedral angle of the tetrahedron, which
equals the dicellular angle of the 8-cell, is $\pi/3$. The volume
integration tells us that such tetrahedra have finite volumes \cite{TH}
though they are non-compact. A single 8-cell includes sixteen vertices
and therefore has sixteen tetrahedra as the boundary after regular
truncations. 

The next step is to find a certain gluing
of appropriate number of regular truncated hyperbolic
8-cell's by identifying the hexahedra, so that the resultant space becomes a
 smooth manifold and the collection of
 the tetrahedra produced by regular truncations form smooth 3-boundaries of the manifold. In mathematical
language, we want to find a discrete subgroup of the isometry group $SO(4,1)$ so that the quotient space
 of $H^4$ by the discrete subgroup
is a manifold. First we would like to try a generalization
to four dimensions of what was illustrated in the simple two dimensional
example and  see whether a manifold can be formed or not. Below is an example of the trial.

\begin{figure}[t]
	\centerline{ \epsfbox{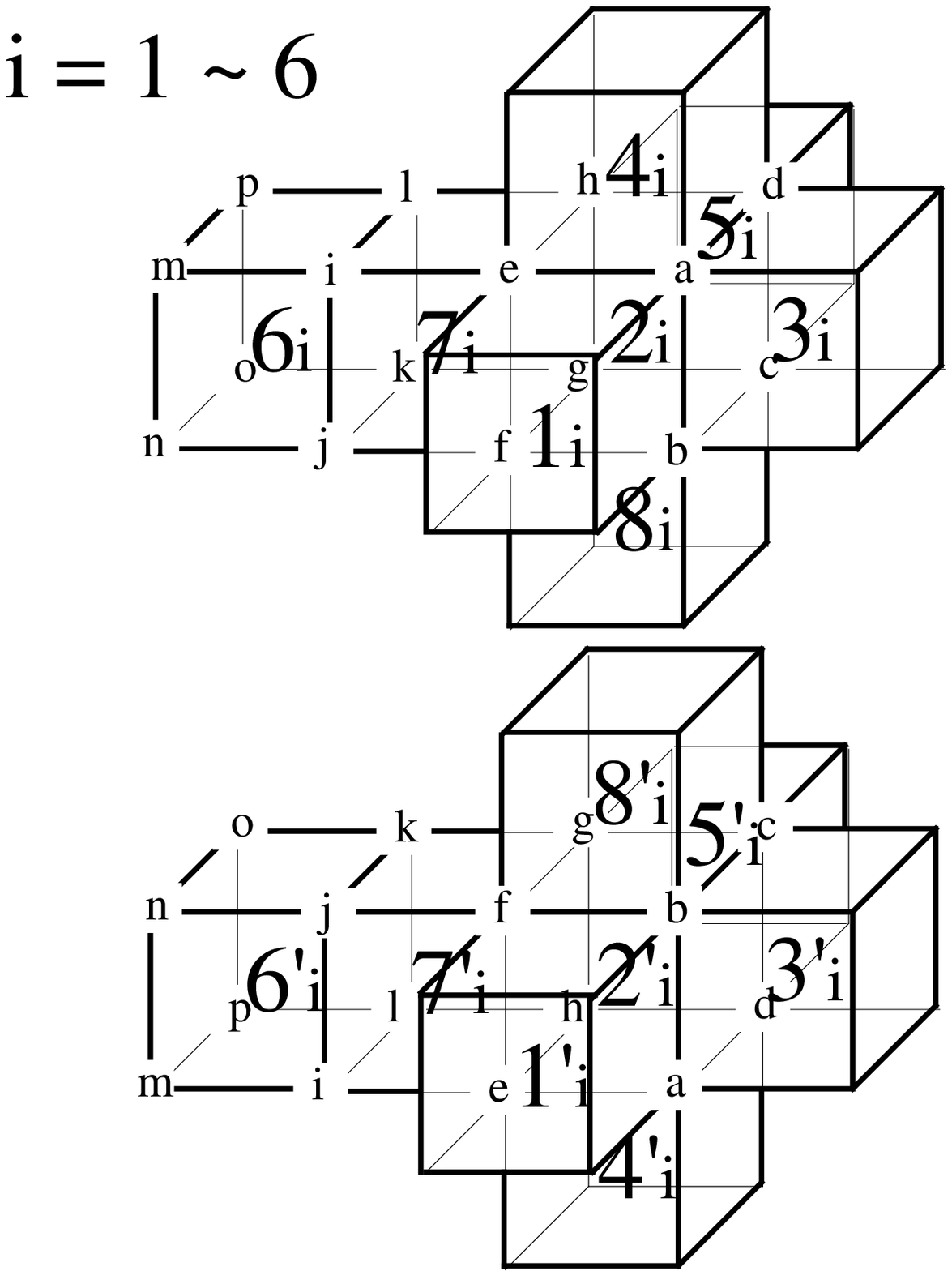}}
        \caption{Two types of 8-cell's. Upper ones are
left-handed while lower ones (with primes) are right-handed. The
corresponding cells with the same number which are primed and unprimed
are identified (for example, $1_1$ and $1'_1$, $4_1$ and $4_3'$ ...).}
\protect\label{fig:const}
\end{figure}

We prepare two regular truncated 8-cell's and put them in the position so 
that they have a reflection
symmetry as depicted in Fig.\ref{fig:const}.
Gluing the hexahedra $X (=1\sim 8)$ and $X' (=1'\sim 8')$ so that all vertices
$(a)\sim(p)$ match, we get a 4-space with sixteen boundary components. 
To make sure smoothness of this 4-space in terms of hyperbolic 
geometry,
it is sufficient to check the smoothness on the boundaries since the
dicellular angles of the 8-cell equal the dihedral angles of the tetrahedra on the boundaries. The inside
of a truncated hyperbolic 8-cell is smooth and regular. We can
nicely glue the two hexahedra in four dimensions because the hexahedra are
totally geodesic. Therefore singular structures possibly appear only on the
faces, edges and/or vertices of each hexahedron. From Fig.\ref{fig:cus2}
we see that the singularity should appear on the boundary tetrahedra
if  any. The gluing of the hexahedra induces the gluing on the
boundary tetrahedra. Fig.\ref{fig:trt}, for example, shows that the
gluing of two tetrahedra corresponding to the vertex $(a)$ both in the
unprimed and primed 8-cell's is determined by the identifications of the
truncated hexahedra around the vertex $(a)$. Since each unprimed hexahedron is identified with
its primed partner, every face of unprimed tetrahedron is glued with
its primed partner so that all vertices match. In this configuration,
the topology of this space composed of the two tetrahedra is $S^3$. 
Nevertheless we have only $2\pi/3$ turning around each edge of the
tetrahedra after gluing, since only two edges of the tetrahedra with the
dihedral angle $\pi/3$ are identified into one edge. This means a
singularity by the deficit angle $2\pi - 2 \times\pi / 3 = 4\pi /
  3$. Therefore we fail to get a smooth manifold by a simple minded
generalization of the 2-dimensional example in the previous section.
  
\begin{figure}[t]
	\centerline{ \epsfbox{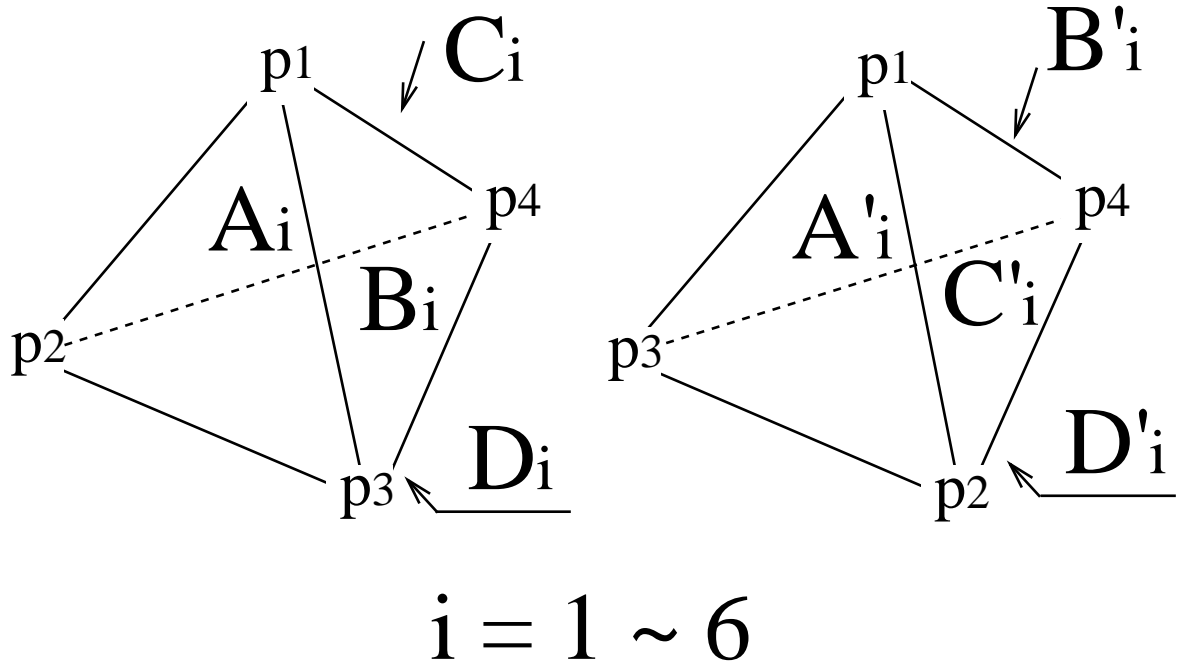}}
        \caption{Six unprimed tetrahedra and six
primed tetrahedra. Twelve tetrahedra constitute $M_B8$.}
\protect\label{fig:fekc}
\end{figure}

However there is a way to improve the construction so that we can get a neat manifold, i.e. without a
deficit angle. We consider a branched
covering space of this singular space. The appropriate branched
covering space can be given by a six-fold cover (twelve tetrahedra) of the
original singular space. The faces and vertices of the twelve
tetrahedra are labeled as Fig.\ref{fig:fekc} ($i=1\sim 6$). The
following pairs of the faces of unprimed and primed tetrahedra are
glued so that all labeled vertices match.
\begin{equation}
\begin{array}{cccc}
        A_1-A'_1 & B_1-B'_3 & C_1-C'_2 & D_1-D'_4 \\
        A_2-A'_2 & B_2-B'_1 & C_2-C'_4 & D_2-D'_3 \\
        A_3-A'_3 & B_3-B'_2 & C_3-C'_5 & D_3-D'_6 \\    
        A_4-A'_4 & B_4-B'_6 & C_4-C'_1 & D_4-D'_5 \\
        A_5-A'_5 & B_5-B'_4 & C_5-C'_6 & D_5-D'_1 \\    
        A_6-A'_6 & B_6-B'_5 & C_6-C'_3 & D_6-D'_2
\end{array}
\label{eqn:gl1} 
\end{equation}
For instance, $A_1$ is matched with $A'_1$. All the vertices $p_1,
p_2, p_3$ of $A_1$ are identified to the vertices $p_1, p_2, p_3$ of
$A'_1$, respectively.

\begin{figure}[t]
	\centerline{ \epsfbox{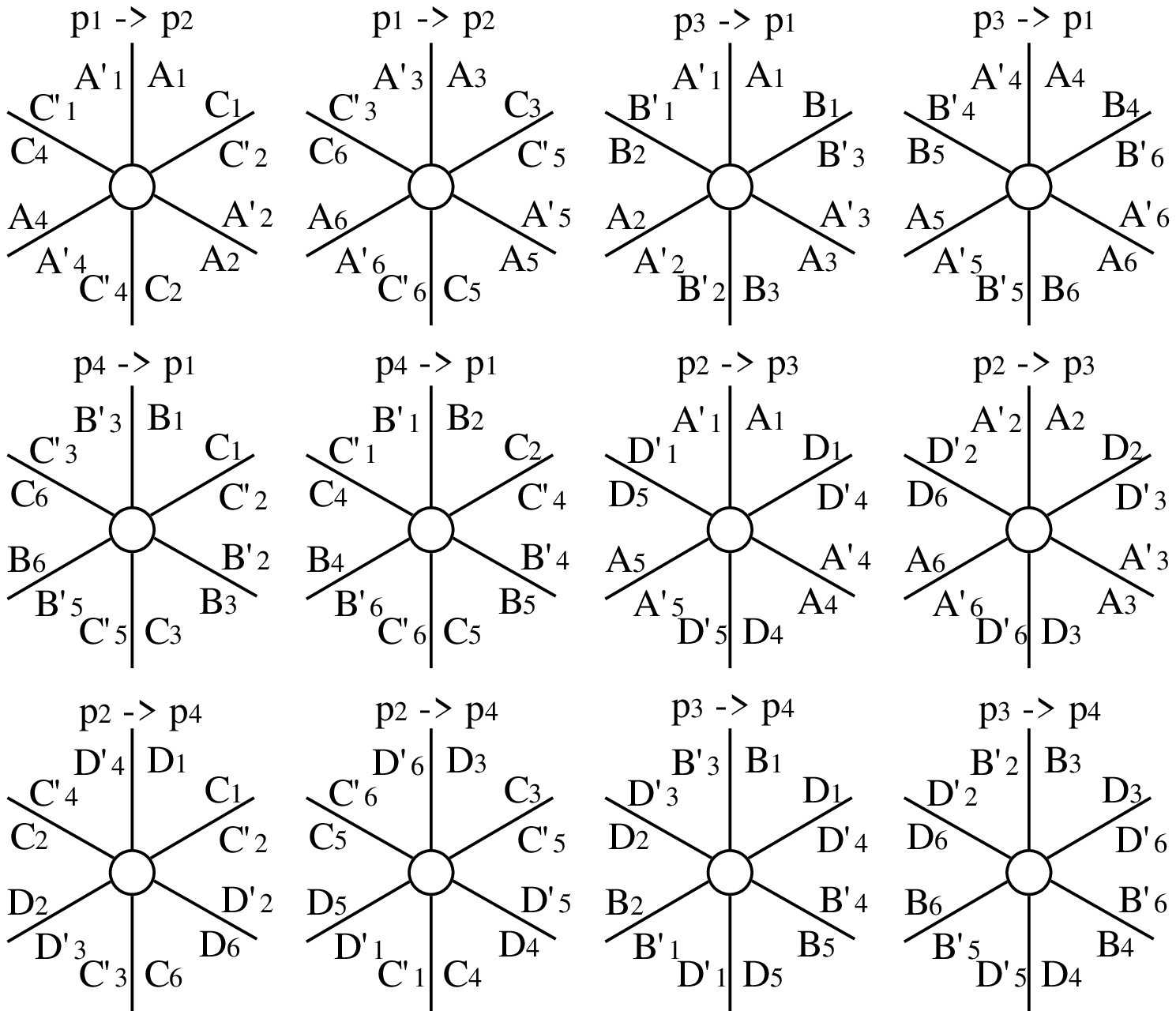}}
        \caption{The consistency check of the manifold which is composed
	of the tetrahedra in Fig.9 with the rule of the equation
	(6). Six dihedral angles meet at the
        edges in Fig.9. Since each dihedral angle is $\pi/3$, 
	there is no deficit angle.}
\protect\label{fig:chk}
\end{figure}

Fig.\ref{fig:chk} shows  consistency of the gluing around every
edge. There are twelve edges in the six-fold covering space after the gluing. There
are six dihedral angles which meet at one edge so that there is no deficit angle since
each dihedral angle is $\pi/3$.

\begin{figure}[t]
	\centerline{ \epsfbox{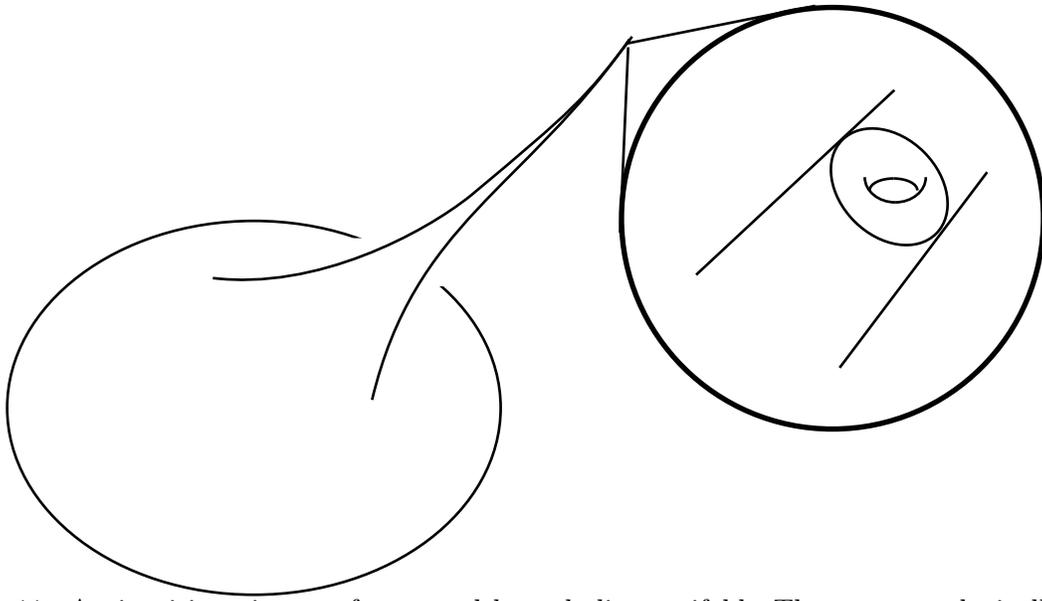}}
        \caption{An intuitive picture of a cusped hyperbolic manifold. The cusp
topologically looks like a torus cross a half-open interval.}
\protect\label{fig:cusp}
\end{figure}

On the other hand, the vertices of the tetrahedra are on the sphere at
infinity $\partial D^3$. By the gluing, these vertices are identified
into four points at infinity, $p_1, p_2, p_3, p_4$. Such points at
infinity are called cusps in the hyperbolic geometry. They are not
singularities of the manifold but open boundaries at
infinity\cite{TH}.  Topologically 
a cusp looks like a torus crossing a half-open interval (see
Fig.\ref{fig:cusp}). The boundary space composed of twelve tetrahedra
is a non-compact smooth manifold, which is called $M_{B8}$ in the
present paper.

The cusp will not cause any serious physical problem because one cannot
observe the infinity of the universe. On the contrary, the existence
of such structures at infinity renders the manifold of primary
importance. It is a known fact in mathematics that there is a family
of almost isometric compact manifolds limiting a cusped
manifold\cite{JW}. It is expected that the limiting cusped manifold shares common characters of the family. Furthermore the cusped manifold
is the simplest one among the family in some sense. Hence, admitting
the cusp to our manifold, we get the following simplest example of
topology changing of spacetime by quantum tunneling.

Now we expect that the branched covering proceeded above can be
straightforwardly extended to the whole of 8-cell's. Since the six-fold cover
of the boundary 3-space produced by the simple minded identification is 
the smooth 
3-manifold $M_{B8}$, the six-fold cover of the 4-space produced by the simple minded identification
 will have smooth boundary
manifolds, $M_{B8}$'s. We prepare six pairs of unprimed and 
primed 8-cell's as Fig.\ref{fig:const} (the pairs are labeled by
$i=1\sim 6$). Every vertex and cell (hexahedron) of the
8-cell's are also labeled in Fig.\ref{fig:const}. All subsequent 
gluings will
be done so that these labeled vertices are matched. We determine the gluing
of hexahedra around each vertex $(a)$ of 8-cell's (cell $1_i\sim$ cell $4_i$, so that they
induce gluing (\ref{eqn:gl1}) on a tetrahedron made by the truncation of the vertex
$(a)$ to form $M_{B8}$.
Fig.\ref{fig:trt} shows the tetrahedron by the 
truncation of each vertex $(a)$. In
this figure, each face of a tetrahedron is labeled by an index of
the cell which the face belongs to, and each vertex of the tetrahedron
is labeled by the same character as that of the nearest vertex of the hexahedron.
From Fig.\ref{fig:trt} and \ref{fig:fekc} we find a correspondence,
$1_i\equiv A_i,\ 4_i\equiv B_i,\ 3_i\equiv C_i,\ 2_i\equiv D_i$. By
(\ref{eqn:gl1}), the gluing of the hexahedra, which constructs $M_{B8}$ from 
the tetrahedra produced by the
truncation of the vertex $(a)$
are
the following:
\begin{equation}
\begin{array}{cccc}
        1_1-1'_1 & 4_1-4'_3 & 3_1-3'_2 & 2_1-2'_4 \\    1_2-1'_2 &
4_2-4'_1 & 3_2-3'_4 & 2_2-2'_3 \\       1_3-1'_3 & 4_3-4'_2 & 3_3-3'_5
& 2_3-2'_6 \\   1_4-1'_4 & 4_4-4'_6 & 3_4-3'_1 & 2_4-2'_5 \\
1_5-1'_5 & 4_5-4'_4 & 3_5-3'_6 & 2_5-2'_1 \\    1_6-1'_6 & 4_6-4'_5 &
3_6-3'_3 & 2_6-2'_2 \ \ .\\ 
\end{array}
\label{eqn:gl2}
\end{equation}

\begin{figure}[t]
	\centerline{ \epsfbox{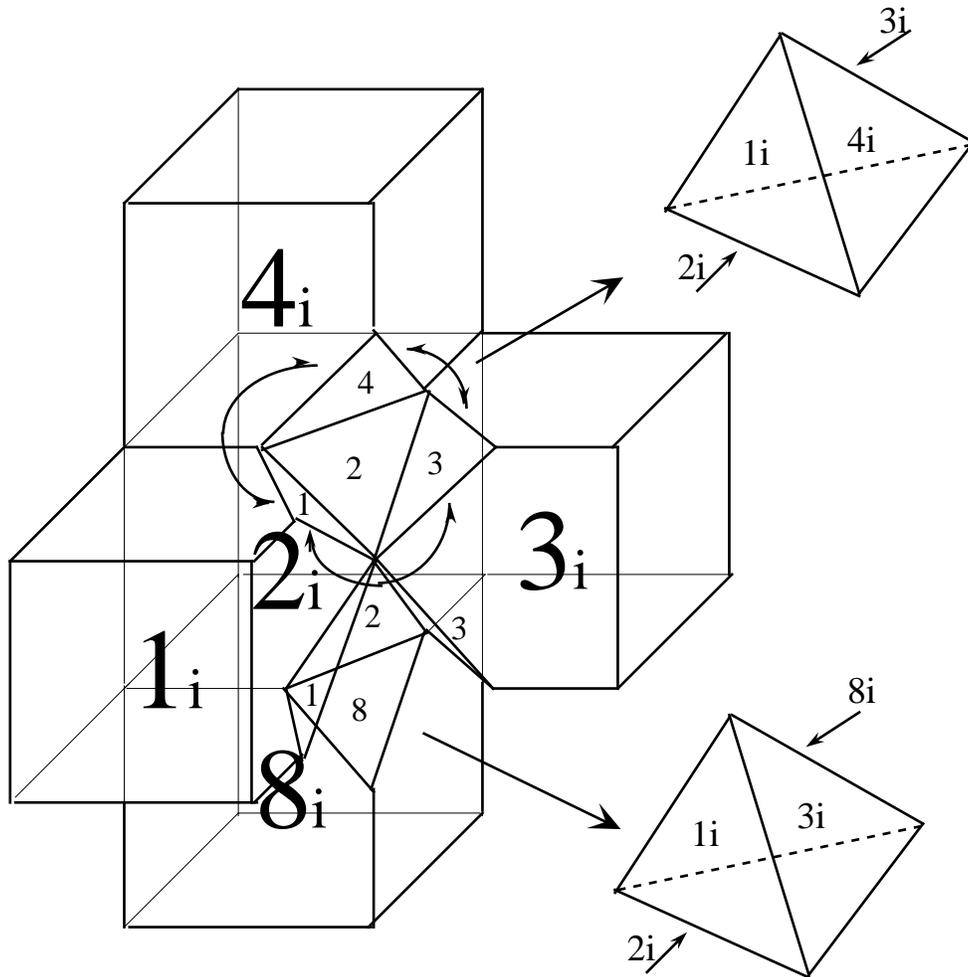}}
        \caption{If we treat $8_i$ as if it were $4_i$, a tetrahedron
from vertex $(a)$ and a tetrahedron from vertex $(b)$ would be
symmetric under the inversion. }
\protect\label{fig:sym}
\end{figure}

It is a non-trivial problem to determine whether it is possible or not
for the tetrahedra produced by the truncations of the other vertices $(b)\sim
(p)$ to form $M_{B8}$'s by 
appropriate choices of gluing of the other cells (cell $5\sim$ cell $8$).
Determining the other gluings as shown below, we can see that two adjacent 
tetrahedra, e.g. formed by the truncations of vertices  $(a)$ and $(b)$
(see Fig.\ref{fig:sym}), are symmetric under the inversion because of
the symmetry of the 8-cell. 

\begin{equation}
\begin{array}{cccc}
        5_1-5'_1 & 8_1-8'_3 & 7_1-7'_2 & 6_1-6'_4 \\    5_2-5'_2 &
8_2-8'_1 & 7_2-7'_4 & 6_2-6'_3 \\       5_3-5'_3 & 8_3-8'_2 & 7_3-7'_5
& 6_3-6'_6 \\   5_4-5'_4 & 8_4-8'_6 & 7_4-7'_1 & 6_4-6'_5 \\
5_5-5'_5 & 8_5-8'_4 & 7_5-7'_6 & 6_5-6'_1 \\    5_6-5'_6 & 8_6-8'_5 &
7_6-7'_3 & 6_6-6'_2
\end{array}\ .
\label{eqn:gl3}
\end{equation}

Since the inversion of one $M_{B8}$ gives another $M_{B8}$, each
group of twelve tetrahedra forms one $M_{B8}$ at each vertex $(b)\sim
(p)$. Therefore the glued twelve 
8-cell's have sixteen $M_{B8}$'s on their boundary. Here a fact should
be noticed that the tetrahedra are orthogonal to the cells (hexahedra)
of the 8-cell, which guarantees that the $M_{B8}$ is smooth at the
points at which the tetrahedra join. Then $M_{B8}$ on the boundary is a totally geodesic
smooth manifold in $H^4$.

Of course these identifications are orientation preserving isometry
transformation because of the reflection symmetry between the unprimed
and primed 8-cell's. The resultant space is orientable.

To check that this 4-space is a complete smooth 4-manifold, we consider
the neighborhood of the faces, edges and vertices. In four dimensions, when
we turn around each face completely, the total angle should be $2\pi$
for consistency. We shall check this consistency on the boundary of it. On the boundary 3-hypersurface, we should check
whether it is $2\pi$ or not around the edges ($\alpha, \beta, \gamma
...$ in Fig.\ref{fig:cus2}) of the tetrahedra. This consistency is
guaranteed by our previous analysis where we have shown that the
boundary consists of several pieces of the manifolds $M_{B8}$ (see
Fig.\ref{fig:chk}).  The 
remaining vertices after the regular truncation ($s, t, u 
...$ in Fig.\ref{fig:cus2}) cause no problem since they form 4-cusps
at infinity.   Hence this space is a complete smooth non-compact
hyperbolic 4-manifold with totally geodesic 3-boundaries. The boundaries are
sixteen $M_{B8}$'s.
%%%%%%%%%%%%%%%%%%%%%%%%%%%%%%%%%%%%%%%%%%%%%%%%%%%%%%%%%%%%%%%%%%%%%%%%%%%%
\subsection{Other Solutions}
As shown previously in the table 1, there are seven kinds of 4-polytopes
admitting regular truncations. Geometrical calculations reveal that
two of them, 24-cell with a dicellular angle $2\pi/5$ and 600-cell
with a dicellular angle $2\pi/3$, are compact and the others are
non-compact after   
regular truncations. We can apply our method for the twelve non-compact
8-cell's to these non-compact cases. By a similar method, we
can successfully get complete smooth hyperbolic 4-manifolds with
totally geodesic 3-boundaries in the two cases. One of them is 
the case of a 16-cell (bounded by sixteen tetrahedra) with a dicellular angle $\pi/2$ and the
other is the case of a 24-cell (bounded by twenty-four octahedra) with a dicellular angle
$\pi/3$. Four 16-cell's form a manifold whose 3-boundaries are
eight $M_{B16}$'s (we consider a two-fold cover of a simple minded identifications),
where an $M_{B16}$ with six cusps is composed of four
octahedra. Similarly, six 24-cell's also constitute a manifold whose
3-boundaries are twenty-four $M_{B24}$'s (we consider a three-fold cover of
a simple minded identifications), where an $M_{B24}$ with eight
cusps is composed of six hexahedra.
 
Here we would like to point out a peculiarity of the hyperbolic
manifold by referring to a mathematical fact. For three or
four dimensional hyperbolic manifold $M_{3,4}$, the fundamental group $\pi_1(M_{3,4})$
determines $M_{3,4}$ uniquely up to an isometry and a choice of
normalizing constants\cite{MP}. This has been known as the Mostow
rigidity. Then it is sometimes sufficient to determine the homology group $H_1$ of
the hyperbolic manifold in 
order to distinguish the manifolds, where $H_1$ is an Abelian group
such that $H_1=\pi_1/[\pi_1,\pi_1]$. To characterize the boundaries
topologically we calculate the corresponding homology groups \cite{NS}, 
\begin{eqnarray}
        H_1(M_{B8}) & = & Z+Z+Z+Z;
        \label{eqn:hg1} \\
        H_1(M_{B16}) & = & Z+Z+Z+Z+Z+Z;
        \label{eqn:hg2} \\
        H_1(M_{B24}) & = & Z+Z+Z+Z+Z+Z+Z+Z.
        \label{eqn:hg3}
\end{eqnarray}
Clearly they are topologically inequivalent. Since the rank of the free 
finite Abelian group part of $H_1$ counts the number of two 
dimensional holes \cite{NS}, $M_{Bn}$ has a more complicated
topological structure than $M_{Bm}$ has if $ n > m$. On the other hand,
the torsion-free property may be considered as a pleasant feature of the
universe because otherwise the universe might
be non-orientable. 
If we had a method which can 
produce a solution whose boundaries have torsion part, much more
solutions could be found.  

Incidentally, it is impossible to construct a solution from 5-cell's 
or 120-cell's in our way. The remaining regular truncated polytopes
are compact. In the compact case, we should check consistency also around the vertices
of the regular truncated polytopes. However it is too complicated for us to work out this consistency check and we need more advanced techniques. This is our project in future \cite{NEXT}. 

%%%%%%%%%%%%%%%%%%%%%%%%%%%%%%%%%%%%%%%%%%%%%%%%%%%%%%%%%%%%%%%%%%%%%%%%%%%%% 
\section{Topology Changing Amplitude and Strong Rigidity}

\begin{figure}[t]
	\centerline{ \epsfbox{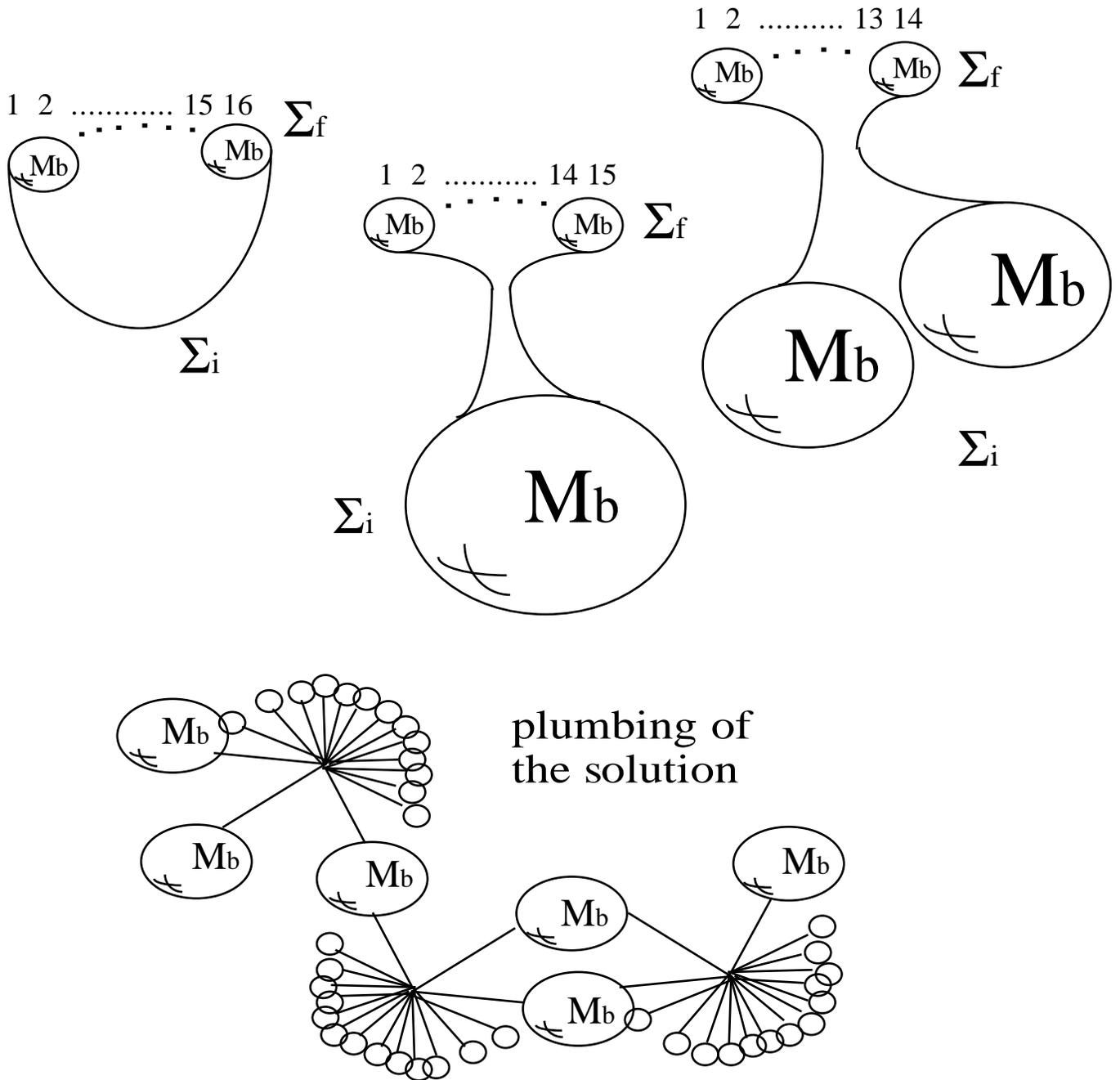}}
        \caption{A Riemannian manifold with sixteen boundaries is regarded as 
        a topology change solution `from nothing to sixteen $M_{B8}$'s',
         `from one $M_{B8}$ to fifteen $M_{B8}$'s' or `from two $M_{B8}$'s to
          fourteen $M_{B8}$'s', and so on. Furthermore, by plumbing of
          the solution we can obtain various types of topology change solutions.}
        \protect\label{fig:plu}
\end{figure}

We have constructed three hyperbolic 4-manifolds with totally geodesic boundaries.
From Gibbons and Hartle\cite{GH}; Fujiwara, Higuchi, Hosoya, Mishima and  
one of the present authors(M. S.) \cite{FHHMS}, these manifolds can be 
regarded as instantons causing topology changes by quantum tunneling.
For example, the manifold of twelve 8-cell's can describe the topology 
changes: `from nothing to sixteen $M_{B8}$'s', `from one $M_{B8}$ to 
fifteen $M_{B8}$'s' or `from two $M_{B8}$'s to fourteen $M_{B8}$'s',
and so on (see Fig.\ref{fig:plu}). It is also worthy of notice that by
plumbing them we can get infinite series of topology changing solutions
as exemplified in  Fig.\ref{fig:plu}. Since each boundary is totally 
geodesic, the gluing is perfect to make a neat manifold when we identify two
boundary components of the same  shape and size. 

We have demonstrated topology changing processes between non-trivial 
spatial topologies in four dimensions by quantum 
tunneling, which cannot be reduced to a lower dimensional subspace. Brill 
constructed a 4-dimensional topology changing solution which, in fact, is
effectively a direct product of a topology changing two dimensional spacetime
and the other two dimensional space\cite{DB}. Of course, the pair creation 
of charged black holes\cite{PC}, in an extended sense, is also a 
process of topology change. The Riemannian manifold for that process 
has a single totally geodesic 3-boundary which can be interpreted as a 
topology change from ``nothing" to the space containing  a pair of 
black holes. However, the present paper gives descriptions of 
processes which include not only the creation of the universe from 
``nothing" but also the change from $n$ $M_B$ to $m$ $M_B$ ($n \neq m$). 
Both of initial and final spatial hypersurfaces ($\Sigma_i$ and 
$\Sigma_f$) have non-trivial topologies.

 Now let us evaluate the
tunneling amplitude for these topology changes. In the context of the
Hawking's Riemannian path integral, the amplitude can be formally
described as 
\begin{equation}
        T(h_i, h_f)=\sum_{M_R}\int{\cal D}g \exp(-S_E[g]),
        \label{eqn:path}
\end{equation}
where $h_i$ and $h_f$ are the 3-dimensional metrics on the initial 
and final spatial hypersurfaces $\Sigma_i$ and
$\Sigma_f$, respectively. $S_E$ is the Euclidean action,
\begin{equation}
        S_E=-{1 \over 16\pi G}\int_{M_R}(R-2\Lambda)\sqrt{g}d^4x
        +{1 \over 8\pi G}\int_{\partial M_R} K \sqrt{h}d^3x \ \ .
        \label{action}
\end{equation}
The path integral is over smooth 4-metric $g$ on the Riemannian manifold 
$M_R$ which has appropriate boundaries $\Sigma_i$ and $\Sigma_f$ by 
assumption. In our cases, $M_R$ is one of the 4-manifolds which have
been constructed in the previous section. Then we can evaluate the
path integral (\ref{eqn:path}) in the WKB approximation for the
topology changing processes. The second term comes 
from the contribution of the boundaries other than $\Sigma_{i,f}$ namely that of
the open boundaries at the cusps. It is easy to see by explicit calculation
that $K$
vanishes at the cusps. Since our solution has a constant negative
curvature $R=4\Lambda<0$, the classical action $\bar{S}_E$ is proportional to the 4-volume of the
spacetime and is
given by  
\begin{equation}
\bar{S}_E={1 \over 8\pi G}{V \over \vert \Lambda \vert},
        \label{act}
\end{equation}
where $V$ is a numerical value of the volume of $M_R$ in the 
case of $\Lambda=-3$. It follows from (\ref{act}) that the WKB 
approximation of the tunneling amplitude is exponentially suppressed for a 
tunneling manifold of a large volume. Then we intuitively expect that the topology
change between more complicated topologies requires a larger volume of tunneling manifold and is
more suppressed provided that the WKB prefactors are of the same order.

Though our manifolds have cusps, their volumes are finite. Following 
Kellerhals \cite{KH}, the hyperbolic volumes of the 4-polytopes that we 
have used are calculated as 
\begin{eqnarray}
        {\rm Volume(truncated\ 8-cell's)}&=&{4\pi^2 \over 3}\\
        {\rm Volume(truncated\ 16-cell's)}&=&{4\pi^2 \over 3}\\
        {\rm Volume(truncated\ 24-cell's)}&=&{20\pi^2 \over 3}
        \label{}
\end{eqnarray}
and the volume of the manifolds are summarized in table 2.

The volume of a constant
curvature space is
given by the Gauss-Bonnet theorem\cite{EGH}, 
\begin{equation}
        \chi(M)={1\over 32\pi^2} \int_M \epsilon_{abcd}
	{\cal R}^{ab}\wedge{\cal R}^{cd}-{1\over 32\pi^2}\int_{\partial M}
	\epsilon_{abcd}(2\theta^{ab}\wedge{\cal R}^{cd}-\frac43\theta^{ab}
	\wedge\theta^c_e\wedge\theta^{ed}),
        \label{volume}
\end{equation}
where ${\cal R}^{ab}$ and $\theta^{ab}$ are the curvature 2-form and the second
fundamental form.
Since the boundary $\partial M$ is totally geodesic, $\theta^{ab}$ vanishes there.
The Euler numbers $\chi(M)$ are combinatorially determined as
\begin{eqnarray}
        \chi({tunneling\ manifold\ consisting\ of\ twelve\ 8-cell's}) & = & 12,
        \label{} \\
        \chi({tunneling\ manifold\ consisting\ of\ four\ 16-cell's}) & = & 4,
        \label{} \\
        \chi({tunneling\ manifold\ consisting\ of\ six\ 24-cell's}) & = & 
        30,
        \label{}
\end{eqnarray}
and agree with (\ref{volume}) and the volumes given in table 2.

Roughly speaking, more polytopes are needed to get a
manifold with a more complicated topological structure. Then we expect 
that the volumes are largely related to the topological structure (just
the Euler number in a constant curvature space). The 
larger volume will imply a more complicated topological structure. Now
let us recall one of the peculiarities of the hyperbolic manifold, Mostow
rigidity. For three or four dimensional hyperbolic manifold $M_{3,4}$,
the fundamental group $\pi_1(M_{3,4})$ determines $M_{3,4}$ uniquely up to  
an isometry and a choice of normalizing constant\cite{MP}. 
Therefore, our tunneling manifolds include no
degrees of freedom of deformation\cite{KHT} as long as the manifold is
hyperbolic. If we include non-zero Weyl curvature, the manifold can become
inhomogeneous and the degrees of freedom of deformation
becomes dynamical. In such a situation, the quantum theory of the
dynamical degrees of freedom have to be developed for a quantum topology
change theory. 

%%%%%%%%%%%%%%%%%%%%%%%%%%%%%%%%%%%%%%%%%%%%%%%%%%%%%%%%
\section{Summary and Discussions}
In the present paper, we have found instantons which describe topology
changing processes by quantum tunneling of spatial hypersurfaces of spacetimes which
are locally anti-de Sitter. 
Here we summarize our results in the table 2.

It may be intuitively expected that the more complicated is the topology of the 
universe which topologically changes, the larger is the volume of the
tunneling manifold. Let us investigate whether this is the case in our 
examples. From table 2, however, we cannot easily draw a conclusion
directly since the number of boundary manifolds are
different. Quantitatively, we can compare a set of three 
topology changing manifolds with $16\times M_{B8}$, a set of six
topology changing manifolds with $8\times M_{B16}$ and a set of two
topology changing manifolds with $24\times M_{B24}$. All the sets have
48 boundaries though they are not arcwise connected. Then the
corresponding manifolds have volumes $48\pi^2$, $32\pi^2$ and
$80\pi^2$, respectively. The result is not just what we
expected. Although the third one is much larger than the other two,
which means that $M_{B24}$ is more unlikely to appear comparing to the other two,
the first one and the second one are comparable. What is
more, the probability for the first one to appear is smaller than that for the
second one. Since $H_1(\Sigma)$ characterizes the topological
structure of $\Sigma$ by the equations (\ref{eqn:hg1}), (\ref{eqn:hg2}) 
and (\ref{eqn:hg3}), the boundary of the first one is simpler than
that of the second one. However, their volumes imply that the topology 
of the first one is more complicated than the topology of the second
one. Nevertheless, we cannot conclude that the tunneling of $M_{B16}$ has
the maximum of probability. There might be a smaller 
manifold describing the topology change of $M_{B8}$'s. If we could find the relation
between the volume of the solution (the Euler number) and the boundary of it,
the relation would explain this.

One might think that our constructions are too restricted. First,
the identification is determined so as to preserve the symmetry 
of the polytope. Second, the resultant polytopes are identical with
each other. From these restrictions, for example, one cannot
consistently identify the polyhedra which bound a single polytope.
Our restrictions made the construction simple. However, it may well be
the case that there are much more solutions which have been ruled out by the 
restrictions. To complete the discussion about the topology change by 
 quantum tunneling, we should relax these restrictions. Then the
structure of the tunneling manifold becomes more involved. 
Constructing such complicated cases will need the aid of computer. 

When we evaluate the topology changing amplitude, the formalism of
Hartle and Hawking has been used. However, the exact no-boundary 
condition in the original formalism of Hartle and Hawking does not
allow the existence of boundary at infinity. In our solutions,
the tunneling manifolds have cusped boundaries at infinity. Since the
cusped boundaries are infinitely small and the manifolds have finite
volumes, we may generalize the formalism of Hartle and Hawking to 
such a case. If we stick to impose the no-boundary condition in a strict sense, we
need compact tunneling manifolds. An investigation in this  
direction is also in progress\cite{NEXT}.

People might be disturbed by the existence of the cusps. 
If we study the effect of ``matter fields'' or fluctuations of the metric 
(``one-loop corrections''), the boundary conditions at the cusps are 
needed. The conditions and the effects will be studied in appropriately 
simplified situation\cite{N2}. Furthermore, some people might insist that the
cusps do not allows one to use Einstein-Hilbert action
and the idea of a smooth manifold in this arbitrary small scale. A calculation in the
(2+1)-dimensional gravity will reveal something about this
as a simplified case. On the other hand, it 
observationally causes no problem since the cusps are at infinity and
we cannot ``see'' them. We see only the pattern of spatial
periodicity\cite{FS}. If we observe the pattern of the spatial periodicity
as the super large scale structure of the universe, we may be able to
determine the topology of our universe and to know whether the
universe has the cusps or not.  

In the case of the topology change in (2+1)-dimensional quantum
tunneling, the rigidity of the hyperbolic manifold is easier to understand. 
While hyperbolic 2-boundaries have moduli parameters as
the dynamical degrees of freedom, the tunneling manifolds do not
allow any deformation corresponding to them, which is consistent with
the rigidity. In four dimensional case, however, the 
situation is different because the 3-boundary is also rigid as well as the
tunneling manifold itself. The dynamical degrees of freedom will appear
only when we allow non-zero Weyl curvature. In such a case the
gravitational degrees of freedom should be considered. As a first step we
can consider the linear perturbation of them. 
If we quantize these degrees of freedom, we expect particles
be created. This might cause a quantum instability of topology changing
solutions if the backreaction to the spacetime is too large.

%%%%%%%%%%%%%%%%%%%%%%%%%%%%%%%%%%%%%%%%
\begin{flushleft}
\Large{\bf Acknowledgments}
\end{flushleft}
We would like to thank Prof. S. Kojima, Prof. M. Sasaki and
 Dr. Higuchi for helpful discussions.
One of the authors (M. S.) thanks the Japan Society for the 
Promotion of Science for financial support. This work was supported in 
part 
by the Japanese Grant-in-Aid for Scientific Research Fund of the Ministry 
of 
Education, Science and Culture.
%%%%%%%%%%%%%%%%%%%%%

%%%%%%%%%%%%%%%%%%%%%%%%%%%%%%%%%%%%%%%%%%
\clearpage


\clearpage

%%%%%%%%%%%%%
\clearpage
\vspace{5mm}

 \centerline{
 \begin{tabular}{c|cccc}
        \hline
        polytope & bounding & dicellular & polyhedra made & 
        solid angle  \\
         & polyhedra & angle & by truncation & 
        around the vertices  \\
        \hline
        5-cell & five tetrahedra & $\pi/3$ & five tetrahedra & 0  \\
        8-cell & eight hexahedra & $\pi/3$ & sixteen tetrahedra & 0  \\
        16-cell & sixteen tetrahedra & $\pi/2$ & eight octahedra & 0  \\
        24-cell & twenty-four octahedra & $2\pi/5$ & twenty-four hexahedra & $4\pi/20$  \\
                & twenty-four octahedra & $\pi/3$ & twenty-four hexahedra & 0  \\
        120-cell & 120 dodecahedra & $\pi/3$ & 600 tetrahedra & 0  \\
        600-cell & 600 tetrahedra & $2\pi/3$ & 120 icosahedra & $4\pi/12$  \\
        \hline
\vspace{5mm}
 \end{tabular}}
Table 1

 The first column is the name of polytopes, which are bounded by
the polyhedra on the second column. The third column gives possible
dicellular angles. After regular truncation, there appear new polyhedra
 shown in the fourth column whose vertices have a solid angle on the
 fifth column. 
\clearpage
\vspace{5mm}
\centerline{
\begin{tabular}{l|l|l|l}
        \hline
        building block & boundary $\Sigma_{i,f}$ & $H_1(\Sigma)$ & the volume of 
        solutions \\
        \hline
        8-cell$\times 12$ & $16\times M_{B8}$ & Z+Z+Z+Z & $12\times 6.8009491$ \\
        16-cell$\times 4$ & $8\times M_{B16}$ & Z+Z+Z+Z+Z+Z & $4\times 8.7730176$  \\
        24-cell$\times 6$ & $24\times M_{B24}$ & Z+Z+Z+Z+Z+Z+Z+Z &
$6\times 59.2029$\\     
       \hline
\vspace{5mm}
\end{tabular}
}
Table 2

The topology changing manifolds that we have constructed. The first
column is the polytope we have used. The resultant manifolds have the boundaries
on the second column whose homology group $H_1$ are shown in the third column.
The volume of solutions are displayed on the fourth column.

\end{document}